\documentclass[12pt]{article}
\usepackage{jheppub}
\usepackage{amsmath,bbm,array,amsfonts,graphicx,wrapfig,lscape,float,slashbox,multirow,longtable,rotating,subfigure,epstopdf,commath}

\newcommand{\be}{\begin{equation}}
\newcommand{\ee}{\end{equation}}
\newcommand{\beq}{\begin{equation}}
\newcommand{\beql}[1]{\begin{equation}\label{#1}}
\newcommand{\eeq}{\end{equation}}
\newcommand{\ba}{\begin{array}}
\newcommand{\ea}{\end{array}}
\newcommand{\bea}{\begin{eqnarray}}
\newcommand{\beal}[1]{\begin{eqnarray}\label{#1}}
\newcommand{\eea}{\end{eqnarray}}
\newcommand{\ben}{\begin{enumerate}}
\newcommand{\een}{\end{enumerate}}
\newcommand{\bean}{\begin{eqnarray*}}
\newcommand{\eean}{\end{eqnarray*}}

\newcommand{\btab}[1]{\begin{tabular}{#1}}
\newcommand{\etab}{\end{tabular}}

\newcommand{\comment}[1]{}

\newcommand{\qed}{\nobreak \ifvmode \relax \else
      \ifdim\lastskip<1.5em \hskip-\lastskip
      \hskip1.5em plus0em minus0.5em \fi \nobreak
      \vrule height0.75em width0.5em depth0.25em\fi}

\def\beqa{\begin{eqnarray}}
\def\eeqa{\end{eqnarray}}
\def\NN{{\cal N}}

\newcolumntype{C}[1]{>{\centering\arraybackslash}m{#1}}

\newcommand{\IZ}{{\mathbb{Z}}}
\def\II{\relax{\rm I\kern-.18em I}}

\def\makeatletter{\catcode`\@=11}% 11:letter
\makeatletter
\def\mathbox#1{\hbox{$\m@th#1$}}%
\def\math@ccstyles#1#2#3#4#5#6#7{{\leavevmode
     \setbox0\mathbox{#6#7}%
     \setbox2\mathbox{#4#5}%
     \dimen@ #3%
     \baselineskip\z@\lineskiplimit#1\lineskip\z@
     \vbox{\ialign{##\crcr
            \hfil \kern #2\box2 \hfil\crcr
            \noalign{\kern\dimen@}%
            \hfil\box0\hfil\crcr}}}}
\def\mathaccstyles{\math@ccstyles\maxdimen}
\def\maththroughstyles{\math@ccstyles{-\maxdimen}}
\def\unity%
{\maththroughstyles{.45\ht0}\z@\displaystyle {\mathchar"006C}\displaystyle 1}
\def\IZ{{\bf Z}}

\def\IT{{\bf T}}

\def\tr{{\rm tr \,}}

\title{On the 3-form formulation of axion potentials from D-brane instantons}
\author[a,b]{Eduardo Garc\'{\i}a-Valdecasas}
\author[a]{, Angel Uranga}

\affiliation[a]{Instituto de F\'isica Te\'orica UAM-CSIC \\
C/ Nicol\'as Cabrera 13-15, Campus de Cantoblanco,  28049 Madrid, Spain}
\affiliation[b]{ Departamento de F\'isica Te\'orica, Universidad Aut\'onoma de Madrid,\\ Campus de Cantoblanco, 28049 Madrid, Spain }
\emailAdd{eduardo.garcia.valdecasas@gmail.com, angel.uranga@uam.es}

\abstract{The study of axion models and quantum corrections to their potential has experienced great progress by phrasing the axion potential in terms of a 3-form field eating up the 2-form field dual to the axion. Such reformulation of the axion potential has been described for axion monodromy models and for axion potentials from non-perturbative gauge dynamics. In this paper we propose a 3-form description of the axion potentials from non-gauge D-brane instantons. Interestingly, the required 3-form field does not arise in the underlying geometry, but rather shows up in the KK compactification in the generalized geometry obtained when the backreaction of the D-brane instanton is taken into account.}

\preprint{
\begin{flushright}IFT-UAM/CSIC-16-049 
\end{flushright} \vspace{-0.9cm}
}

\begin{document}

\maketitle

%==================================================%
%==================================================%

%===============================================================================
\section{Introduction and main results} \label{sec:intro}
%===============================================================================

Axions have become an essential template to describe physics of scalar fields whose potential enjoys special protection properties due to an underlying symmetry principle. Naively, the symmetry corresponds to the perturbative global symmetry shifting the value of the scalar field, which is violated by non-perturbative effects, as originally proposed for the QCD axion \cite{Peccei:1977hh}. However, it has recently become clear that the most fundamental symmetry structure is that of the dual 2-form. Contributions to the axion potential which spoil the shift symmetry must arise from the existence of a 3-form which eats up the dual 2-form to make it (and so the dual axion) massive. The gauge symmetry of the 3-form constrains the form of these contributions in an advantageous way for many phenomenological applications. 
The description of the axions in terms of forms and their duals has also been key to the use of the weak gravity conjecture \cite{ArkaniHamed:2006dz} to constrain transplanckian axion model building \cite{delaFuente:2014aca,Rudelius:2015xta,Montero:2015ofa,Brown:2015iha,Bachlechner:2015qja,Hebecker:2015rya,Brown:2015lia,Junghans:2015hba,
Heidenreich:2015wga,Bielleman:2015ina,Palti:2015xra,Heidenreich:2015nta,Ibanez:2015fcv,Hebecker:2015zss,
Hebecker:2015tzo,Conlon:2016aea,Baume:2016psm,Heidenreich:2016jrl}.

This formulation has been well understood: 

$\bullet$ For the QCD axion, in \cite{Dvali:2005an,Dvali:2005ws,Dvali:2013cpa}, where the 3-form is actually the Chern-Simons composite 3-form built out of the QCD gauge fields. 

$\bullet$ In string compactifications producing axion monodromy \cite{Silverstein:2008sg,McAllister:2008hb}, as described in \cite{Marchesano:2014mla} connecting it to the earlier description in \cite{Kaloper:2008fb, Kaloper:2011jz}  \footnote{For other works related to axion monodromy, see \cite{Berg:2009tg, Ibanez:2014zsa,Palti:2014kza,Blumenhagen:2014gta,Ibanez:2014kia,Hebecker:2014eua,Kaloper:2014zba,Arends:2014qca,Franco:2014hsa,Blumenhagen:2014nba,Hebecker:2014kva,Ibanez:2014swa,Garcia-Etxebarria:2014wla,Blumenhagen:2015kja,Retolaza:2015sta,Escobar:2015fda,Bielleman:2015lka,Blumenhagen:2015xpa,Ibanez:2015fcv}.}. In these cases, the 3-form is a fundamental field, and its couplings arise from different sources, ranging from Chern-Simons couplings to fluxes in the 10d action \cite{Marchesano:2014mla,McAllister:2014mpa} (see also \cite{BerasaluceGonzalez:2012zn}), torsion homology \cite{Marchesano:2014mla} (see also \cite{Camara:2011jg}) or topological brane-bulk couplings \cite{Retolaza:2015sta}.

The two above phenomena, in particular the presence of fluxes and non-perturbative effects on D-brane gauge sectors, play an important role in several scenarios of moduli stabilization (and thus of their axion components), along the lines in \cite{Kachru:2003aw}. Actually, the gauge non-perturbative effects can be described in string theory as particular cases of D-brane instanton effects wrapping the same cycle as the gauge D-branes in the compact space. In general, in string theory there are other non-perturbative effects from D-brane instantons not wrapping such cycles (sometimes dubbed stringy or exotic D-brane instantons \cite{Blumenhagen:2006xt,Ibanez:2006da,Florea:2006si}, see \cite{Blumenhagen:2009qh,Ibanez:2012zz} for reviews), and contributing to the stabilization of axions as well. It is therefore natural to wonder about the 3-form description of these latter effects. Interestingly, there is no known description of this kind: since there is no gauge group associated to the cycles, we cannot use any composite Chern-Simons 3-form; on the other hand, for e.g. a D3-brane instanton on a 4-cycle, there is not any obvious corresponding harmonic form able to produce a 3-form in the 4d theory upon compactification.

In this paper we solve this question and provide the 3-form description for the stabilization of an axion by non-gauge D-brane instanton effects. The key idea is to notice that the stabilizaton occurs when the non-perturbative effect is included in the theory, so it is only then that we can hope to find a suitable 3-form. Therefore, the internal form supporting the 4d 3-form must arise only in the geometry backreacted by the presence of the D-brane instanton, in the sense discussed in \cite{Koerber:2007xk,Koerber:2008sx}. In general, these correspond to generalized geometries, so the corresponding form need not be harmonic with respect to the underlying CY metric, rather it corresponds to (a piece of) a generalized calibration.

We study this in the particular example of D3-brane instantons on 4-cycles, but the lesson is general (as expected from T-duality / mirror symmetry). Also, we show that the picture is compatible with D-brane instantons corresponding to gauge non-perturbative effects.

The paper is organized as follows. In Section \ref{sec:review} we review the 3-form description of axion stabilization and its interplay with axion monodromy and non-perturbative gauge dynamics. In Section \ref{sec:themeat} we provide the  3-form description of axion potentials induced by non-gauge D-brane instantons: after posing the question in section \ref{sec:puzzle}, we review the D3-brane instanton backreacted geometry in section \ref{sec:backreaction}, and obtain the 4d 3-form and its couplings in section \ref{sec:the3form}. A simple example is displayed in section \ref{sec:example}. 
Section \ref{sec:general} describes the generalization, in particular the mirror picture of D2-brane instantons in type IIA compactifications.
In section \ref{sec:gauge} we discuss the case of gauge D-brane instantons. Finally, Section \ref{sec:conclusions} contains our final remarks.

\section{Review of 3-forms and monodromy.}
\label{sec:review}

Consider an axion $\phi$, regarded just as a scalar taking values in a circle (i.e. with discrete periodicity\footnote{For simplicity we set the axion decay constant to $f_\phi=1$.} $2\pi $) and with an (approximate) shift symmetry. In many applications one is interested in generating a non-trivial potential for this axion, violating precisely this shift symmetry. For simplicity we consider the potential expanded at quadratic order around a minimum, as for instance arises in moduli stabilization; the general picture is however more general. Hence we  have the lagrangian
\beqa
|d\phi|^2\, +\, \mu^2\, \phi^2
\label{massive-axion}
\eeqa
A potential of this kind is naively not compatible with the axion periodicity, but may be made so by including multivalued branches for the potential, thus inducing axion monodromy. The description of the periodicity properties are automatically implemented by using a dual formulation, in terms of a 3-form $c_3$ eating up the 4d dual  2-form field $b_2$ (defined by $*\, db_2=d\phi$, in 4d). This is described by the following lagrangian\footnote{We allow for a $\IZ_n$ discrete gauge symmetry, see \cite{BerasaluceGonzalez:2012zn,Berasaluce-Gonzalez:2013bba,Marchesano:2014mla} for discussions of such system and the corresponding $\IZ_n$ charged domain walls.}
\beqa
|db_2+nc_3|^2 +|F_4|^2
\eeqa
where $F_4=dc_3$. This theory has the gauge invariance
\beqa
c_3\to c_3+d\Lambda_2\quad ,\quad b_2\to b_2-n\Lambda_2
\label{gauge-inv}
\eeqa
As emphasized in \cite{Marchesano:2014mla}, it is this gauge symmetry that protects the flatness of the axion potential against uncontrolled corrections even in transplanckian field ranges. Dualizing the 2-form back into the axion $\phi$, we obtain the
Kaloper-Sorbo description of axion monodromy \cite{Kaloper:2008fb, Kaloper:2011jz, Dubovsky:2011tu, Kaloper:2014zba}
\beqa
|d\phi|^2\, +\, n\phi\, F_4\, +\, |F_4|^2
\eeqa
Integrating out the non-dynamical 4-form field strength $F_4$ one recovers a potential with the structure (\ref{massive-axion}).

The idea can be generalized to other potentials by considering corrections, which due to the gauge invariance, (\ref{gauge-inv}) must be in terms of higher powers in $F_4$. This leads for instance to the flattening in \cite{Dong:2010in}.

\medskip

At this point we would like to emphasize an important clarification: the description of the massive axion in terms of a 3-form eating up the dual 2-form, or in terms of a coupling between (a function of) the axion and the 4-form field strength, is not necessarily a signature of axion monodromy, but rather of the existence of a non-trivial axion potential. Axion monodromy arises in cases when the potential does not naively satisfy the axion periodicity. On the other hand,  the 3-form description should also hold for non-monodromic potentials (i.e. single-valued and consistent with the axion periodicity). Our main interest in this note is indeed the 3-form description of single instanton non-perturbative potentials, which indeed are periodic in the axion.

\medskip

Before entering this discussion, we conclude this review with a brief recap of the 3-form description of axion potentials induced by non-perturbative gauge dynamics.  This has been studied in the context of QCD axions in \cite{Dvali:2005an,Dvali:2005ws,Dvali:2013cpa}, but we recast it for pure $\NN=1$ $SU(n)$ SYM, which will be useful later on.
The axion belongs to a chiral multiplet $\phi={\rm Im}\, T$, and we have a coupling
\beqa
S_{\rm SYM} \,=\, \int d^2\theta d^4x \, T\, W_\alpha W^\alpha\,\sim\, \int_{4d} \, \phi\, \tr F^2\, +\ldots
\label{susceptibility}
\eeqa
The non-perturbative dynamics produces a gaugino condensate superpotential
\beqa
W\, =\, \omega^k \Lambda^3\, =\, \omega^k\, e^{-\frac T{n}}
\eeqa
where $\omega=e^{2\pi i /n}$ and $\Lambda=\exp\big(-\frac T{3n}\big)$ is the SYM dynamical scale, regarded as a function of $T$.

The theory has $n$ gapped vacua, which differ by the value of $\langle \phi\rangle$ (the phase of the gaugino condensate), in the sense that a shift $\phi\to \phi +2\pi$ changes the $k^{th}$ vacuum to the $(k+1)^{th}$. Therefore the potential has a periodicity of $\phi\sim \phi + 2\pi n$. On the other hand, the axion should actually have a periodicity of $2\pi$; in fact, this is ensured because there is a 3-form description of the above axion potential. This follows from realizing that the coupling (\ref{susceptibility}) describes a coupling to a 4-form $\phi F_4$ structure, with $F_4=\tr F^2$. Namely, the 3-form is the composite Chern-Simons 3-form $c_3=\omega_3^{\rm CS}$ (defined by $d\omega_3^{\rm CS}=\tr F^2$).

In other words, although the potential is periodic with a period $2\pi n$, the periodicity of $2\pi$ is achieved via a monodromic structure, but with a finite number $n$ of branches. This is clearly associated with the $\IZ_n$ discrete symmetry among the $n$ vacua of $\NN=1$ SYM. The structure of domain walls among vacua is naturally described in terms of the (composite) 3-form description \cite{Dvali:2005an,Dvali:2005ws,Dvali:2013cpa}.

\section{3-forms from D-brane instanton backreaction}
\label{sec:themeat}

\subsection{The puzzle}
\label{sec:puzzle}

In string theory, there are non-perturbative contributions to the superpotential beyond the above non-perturbative gauge dynamics contributions. These come from, for instance, euclidean D-brane instantons which do not correspond to gauge theory instantons. A prototypical example is provided by D3-branes instantons wrapped on 4-cycles in CY compactifications\footnote{In general, these may carry world-volume fluxes, but for simplicity we will restrict to the case of trivial gauge backgrounds.}, such as those used in moduli stabilization in \cite{Kachru:2003aw}. In our discussions we will assume the instantons to indeed produce superpotential terms; it would be interesting, but beyond our scope, to develop an understanding of a dual description for instanton effects generating higher F-terms due to extra fermion zero modes \cite{Beasley:2004ys,GarciaEtxebarria:2008pi}. Denoting by $T$ the complex modulus associated to the wrapped 4-cycle, we have a superpotential
\beqa
W_{\rm np}\, =\, A\, e^{-T}
\label{npsupo}
\eeqa
where $A$ is a prefactor depending on complex structure moduli (and possibly also on open string moduli), whose detailed structure is not essential for our present purposes.

We expect the axion potential induced by these instantons to admit a description in terms of a 3-form eating up the dual 2-form. On the other hand, there is no obvious candidate for such a 3-form in the 4d CY compactification: the only available RR fields are even-degree gauge potentials, so the 3-form should arise from the RR 6-form integrated over a 3-cycle. However, there is no natural pairing between a 3-cycle and a 4-cycle in a CY so as to support the topological coupling $\phi F_4$ ultimately responsible for the axion potential. Therefore there is no natural candidate for the 3-form coupling to the axion to reproduce its potential.

The solution to this problem is to follow the intuition gained in the discussion of the SYM superpotential. In that case, the 3-form arises only when the existence of non-perturbative sectors in the gauge theory are taken into account; namely, the fact that $F_4=\tr F^2$ implies that  the presence of the instantons is built-in in the coupling $\phi F_4$. The implementation of a similar concept for non-gauge D-brane instantons requires proposing that the 3-form describing the axion stabilization should be looked for not in the original CY geometry, but rather in the geometry perturbed by the gravitational backreaction of the D-brane instantons. This perturbation of the geometry has been studied in the literature in  \cite{Koerber:2007xk,Koerber:2008sx} by exploiting the technology of generalized geometry.

\subsection{D-brane instanton backreaction}
\label{sec:backreaction}

In this section we review ideas in \cite{Koerber:2007xk,Koerber:2008sx} to describe the backreaction of D-brane instantons in terms of generalized geometry. The uninterested reader may wish to jump to section \ref{sec:the3form}.

The effect of D-brane instantons can be encoded in the underlying CY geometry by means of a deformation turning the $SU(3)$ holonomy into (in general) an $SU(3)\times SU(3)$ structure, associated to the existence of two spinors (not covariantly constant due to the deformation) corresponding to a 4d $\NN=1$ supersymmetry (possibly in AdS). Focusing already in the type IIB case, the two spinors are written
\beqa
\epsilon_1=\zeta_+\otimes \eta_+^{(1)}+\zeta_-\otimes \eta_-^{(1)}\quad , \quad
\epsilon_2=\zeta_+\otimes \eta_+^{(2)}+\zeta_-\otimes \eta_-^{(2)}
\eeqa
here $\zeta_+$ and $\eta_+$ are complex conjugate of $\zeta_-$, $\eta_-$, and $\zeta_+$ is the 4d spinor specifying the $\NN=1$ supersymmetry, and satisfying $\nabla_\mu \zeta_-=-\frac 12 W_0 \gamma_\mu \zeta_+$, where $W_0$ is the superpotential at the AdS minimum (and for the Minkowski case we just have $W_0=0$).

The spinors $\eta^{(1,2)}$ can be used to define two polyforms, 
\beqa
\Psi_{\pm}=-\frac{i}{||\eta^{(1)}||^2}\, \sum_l \frac{1}{l!} \,\eta_{\pm}^{(2)}{}^\dagger\gamma_{m_1\ldots m_l} \eta_+^{(1)}\, dy^{m_l}\wedge \ldots \wedge dy^{m_1}
\eeqa
Noticing the chirality of the spinors in the sandwich, the polyform $\Psi_+$ contains even degree forms and $\Psi_-$ contains odd degree forms.
A common alternative notation is (for type IIB) $\Psi_1=\Psi_+$ and $\Psi_2=\Psi_-$.

The familiar case of $SU(3)$ structure corresponds to $\eta^{(2)}\sim \eta^{(1)}$ and leads to  $\Psi_+\sim e^{iJ}$ and $\Psi_-\sim \Omega$. For $SU(3)$ holonomy the spinors are covariantly constant and the polyforms are closed.

The compactification ansatz is
\beqa
ds^2\, =\, e^{2A(y)}g_{\mu\nu}(x) \,dx^\mu dx^\nu \, +\, h_{mn}(y)\, dy^mdy^n
\eeqa
The 10d fields can be organized in complex quantities, in agreement with the 4d susy structure. One holomorphic quantity is
\beqa
{\cal Z}\equiv e^{3A-\Phi} \Psi_2
\eeqa
where $\Phi$ is the 10d dilaton (not to be confused with the 4d axion).
This is motivated because it provides the calibration for BPS domain wall D-branes (notice that in IIB, $\Psi_2$ calibrates odd-dimensional cycles). In other words, the tension of a  D-brane BPS domain wall is obtained by integrating the above form over the wrapped cycle. For instance, for standard CY compactifications, a D5-brane on a supersymmetric 3-cycle $\Pi$ provides a 4d BPS domain wall, whose tension is given by $\int_\Pi {\cal Z}_{(3)}\sim \int_{\Pi}\Omega$, where the subindex $(3)$ denotes the 3-form part of the polyform

The second quantity is 
\beqa
{\cal T}\equiv e^{-\Phi } {\rm Re}\, \Psi_1+i\Delta C \label{eq:TPolyform}
\eeqa
where $\Delta C$ describes the RR backgrounds not encoded in the background fluxes ${\bar F}$, i.e. the RR fluxes are split as $F={\bar F}+d\Delta C$. The above complexification is motivated by the generalized calibration of BPS D-brane instantons (notice that in IIB, $\Psi_1$ calibrates even-dimensional cycles). For instance, for standard CY compactifications, a D3-brane wrapped on a holomorphic 4-cycle $\Sigma$ provides a 4d BPS instanton whose action is given by $\int_\Sigma {\cal T}_{(4)}\sim \int_{\Sigma} e^{iJ}\sim \int_\Sigma J\wedge J$.

\subsection{The 3-form and its coupling}
\label{sec:the3form}

From the supersymmetry conditions recast in terms of the 10d version of the 4d fields ${\cal Z}$, ${\cal T}$, one can show that, in a weak coupling expansion,
we have
\beqa
d_H{\cal Z}\,=\, \frac{2i}{n}W_{\rm np}\,\delta_2(\Sigma)
\label{magic}
\eeqa
where $d_H=d+H\wedge$, so the 2-form term above is just $d_H{\cal Z}_{(1)}=d{\cal Z}_{(1)}$. This encodes the backreaction of the instanton effect on the 10d geometry in terms of the appearance of a 1-form component ${\cal Z}_{(1)}$ of ${\cal Z}$, which was absent in the CY geometry. 

This equation defines an special 1-form $\alpha_1\equiv {\cal Z}_{(1)}$. It is associated to the globally defined supersymmetric spinor, in the presence of the non-perturbative correction to the geometry. Its structure can be obtained by integrating the above equation. For the particular case of a D3-brane instanton ($n=1$) on a holomorphic 4-cycle defined by the equation $f=0$, one obtains \cite{Koerber:2007xk}
\beqa
{\cal Z}_{(1)}\, \sim\, df\, {\tilde W}_{\rm np}
\eeqa
where the tilded superpotential has the dependence on open string degrees of freedom removed.

One intuitive way to understand the above expression is to notice that,  in a theory containing gauge D3-branes (i.e. D3-branes sitting at a point in the internal space), the 4d superpotential as a function of the D3-brane position is obtained by considering a 1-chain $L$ joining two different points in the CY and integrating ${\cal Z}_{(1)}$ over it. This follows because ${\cal Z}_{(1)}$ is the calibrating form for a D3-brane wrapped on the 1-chain $L$, which defines a domain wall interpolating between the two configurations of the D3-branes at the two (end)points. We thus have
\beqa
\Delta W\, \sim\, \int_L {\cal Z}_{(1)}\, =\, f\,  {\tilde W}_{\rm np}
\eeqa
where we regard $f$ as the D3-brane position in the direction normal to the instanton 4-cycle. The result therefore reproduces the familiar dependence on open string moduli, microscopically associated to Ganor strings \cite{Ganor:1996pe} (see also \cite{Baumann:2006th}).

In the following we recast (\ref{magic}) as $d\alpha_1=\beta_2$, 
and we use both $\alpha_1$ and $\beta_2$ as internal profiles for the KK reduction of higher-dimensional form fields in the backreacted geometry.
This is similar to the non-harmonic forms used in KK reduction of massive $U(1)$'s, and studied in \cite{Camara:2011jg} in compactification spaces with torsion (co)homology.

 We may use these forms to perform the KK reduction of the 10d RR 4-form $C_4$ as
\beqa
C_4\,=\, \alpha_1(y)\wedge c_3(x)\, +\, \beta_2(y)\wedge b_2(x)\,+\ldots
\label{expand-in-3form}
\eeqa
This produces a 3-form in 4d spacetime, naturally associated to the non-perturbative effect, and a 2-form, dual of $\phi$. Moreover, is it clear that the 3-form is eating up the 2-form, by noticing that the field strength $F_5$ has a term
\beqa
F_5= (1 + *_{10d}) \left( \beta_2 \wedge (c_3 +db_2)- \alpha_1 \wedge F_4 \right)
\label{5-form}
\eeqa
where $ *_{10d}$ is added to take the self-duality of $F_5$ into account. This clearly has the gauge invariance
\beqa
c_3\to c_3+d\Lambda_2 \quad , \quad b_2\to b_2-\Lambda_2
\eeqa
This implies that the 3-form is eating up the 2-form to become massive, and correspondingly provides a dual description of the axion $\phi$ becoming massive, as in Section \ref{sec:review}.

It is also straightforward to show that this 3-form has a Kaloper-Sorbo coupling to the axion. We focus just on the leading term $\phi F_4$, where $\phi=\int_\Sigma C_4$, and $F_4=dc_3$. We simply massage the kinetic term of the (self-dual) 4-form and focus on the components in (\ref{5-form}) as follows
\beqa
\int_{10d} F_5\wedge * F_5\, =\, -\int_{10d} C_4\wedge d * F_5 \, \rightarrow \, \int_{10d} C_4\wedge \beta_2 \wedge F_4\, =\, \int_{4d} \phi\, F_4 \nonumber
\eeqa
where we used $\beta_2\sim \delta_2(\Sigma)$.

\subsection{Some toroidal examples}
\label{sec:example}

To flesh out this somewhat abstract description, let us now consider a toroidal compactification  $M_4 \times \IT^6$, where for simplicity we take a  factorizable, $\IT^6 = \IT^2 \times \IT^2 \times \IT^2$ with local complex coordinates be $z_1, z_2, z_3$. 
Let us study the backreaction caused by a instantonic D3-brane wrapping the 4-cycle $\Sigma_4$ defined by $z_3=0$. 
Using the general formulas in \cite{Koerber:2007xk,Dymarsky:2010mf}, the complex structure ${\cal Z}=\Omega$ gets corrected, becoming a generalized complex structure with a 1-form piece
\begin{equation}
Z_{(1)}\sim  e^{-T} \, \, d z_3, \label{eq:ComplexDeformationD3}
\end{equation}
This is the 1-form to be used to produce the 4d 3-form upon compactification of the 10d RR 4-form. 

Notice that it actually corresponds to $dz_3$, a harmonic 1-form already present in the underlying toroidal geometry. Therefore there seems to be essentially no new geometric structure associated to the  backreacted geometry, namely, no axion potential due to the instanton effect. This feature is clearly related to the existence of extra harmonic forms in the $\IT^6$ geometry, which are not present in generic CYs. However, it nicely dovetails the expectation that D3-brane instantons in toroidal geometries have additional fermion zero modes, and do not  produce non-perturbative superpotentials for the corresponding moduli. 

\medskip

In order to actually get non-trivial structure, we can consider orbifolds which remove the extra harmonic forms, and produce genuine CY geometries. Consider for instance $\IT^6/(\IZ_2\times\IZ_2)$, where the generators of the orbifold group act as $\theta:(z_1,z_2,z_3)\to (-z_1,z_2,-z_3)$, and $\omega: (z_1,z_2,z_3)\to (z_1,-z_2,-z_3)$. To describe the quotient, we introduce local coordinates\footnote{A global construction is easily produced by using Weierstrass equations for the 2-tori, but we will not need this extra complication.} by building orbifold invariants, $u_i=z_i^2$, $t=z_1z_2z_3$, subject to $u_1u_2u_3=t^2$. The instanton wrapped on the second and third torus is defined by $f\equiv u_3=0$, so $f=u_3$, and we have
\beqa
{\cal Z}_{(1)}\sim du_3\sim z_3\,dz_3
\eeqa
It is now clear that the 1-form supporting the 3-form in the compactification of the 10d 4-form is non-harmonic with respect to the original CY geometry. 

\subsection{Generalization}
\label{sec:general}

Although we have focused on the case of axions with potential arising from D3-brane instantons on 4-cycles, the ideas hold for general RR axions associated to other cycles, and with potentials arising from the corresponding wrapped D-brane instantons. In order to illustrate this, we consider the mirror configuration of type IIA with axions arising from the RR 3-form on a 3-cycle, stabilized by D2-brane instantons.

Let us thus study the  mirror dual to the configuration of D3-branes on a 4-cycle, in the setup of a general CY (alternatively the main ideas can already be illustrated in the toroidal examples in section \ref{sec:example}). Consider the CY in the large complex structure limit, where it can be regarded as a $\IT^3$ (parametrized by coordinates $y_i$, $1,2,3,$) fibered over a 3d base, with local coordinates $x_i$, $i=1,2,3$. The complex coordinates are locally $z_i = x_i + i y_i$, and we consider a holomorphic 4-cycle given locally by $z_3=0$, i.e. spanning  $x_1$, $y_1$, $x_2$, $y_2$.

The mirror dual can be obtained by applying  three T-dualities \cite{Strominger:1996it}, along the coordinates $y_i$. The D3-brane instanton thus turns into a D2-brane wrapped on the 3-cycles locally spanned by $x_1$, $x_2$, $y_3'$ (with the prime denoting the T-dual coordinate).  One further sees that the complex structure deformation $Z_{(1)} \sim d f=d x_3 + i d y_3$ gives rise to a polyform 
\begin{equation}
\delta \mathcal{T} = \mathcal{T}_{(2)} + \mathcal{T}_{(4)}, \label{eq:TDeformations}
\end{equation}
This follows from the fact that ${\cal Z}_{(1)}$ is eventually use to expand the RR forms and obtain the 4d 3-form. Hence, these are the 2- and 4-form components of (\ref{eq:TPolyform}) produced by the backreaction of the D2-instanton, as we argue  later on. Before that, let us conclude that the expansion of the RR polyform ${\cal C}=C_1+C_3+C_5+C_7$ along $\delta {\cal T}$ produces the 4d 3-form as follows
\beqa
{\cal C}\, =\, \delta {\cal T} \wedge c_3 \; \rightarrow\; C_7={\cal T}_{(4)}\wedge c_3\; , \;  C_5={\cal T}_{(2)}\wedge c_3
\eeqa
where the $c_3$ in the last two expansions is understood to be the same 4d 3-form.

\medskip

Let us finish by arguing further that the above $\delta{\cal T}$ is indeed the backreaction corresponding to the D2-brane instanton. In the original picture in section  \ref{sec:the3form} we considered the superpotential in the theory in the presence of a gauge D3-brane, given by the integral of the calibrating form over a 1-chain.
Under the mirror transformation we must consider the theory in the presence of a gauge D6-brane wrapped on the 3-cycle $\Pi_3$ (i.e. the mirror $\IT^3$ fiber, spanned by $y_i'$, $i=1,2,3$). The superpotential is given by the integral of the calibrating form ${\cal T}$ over a (generalized) 4-chain $\Sigma$ interpolating between two (possibly generalized) 3-cycles (see e.g. \cite{Marchesano:2014iea})
\begin{equation}
\Delta W_{D6} = \int_{\Sigma} \delta \mathcal{T} 
\end{equation}
We are interested in the superpotential as a function of one D6-brane complex modulus, given by one deformation of the special lagrangian 3-cycle, and the corresponding Wilson line along one 1-cycle in $\Pi_3$. The actual components turned on in this $\delta {\cal T}$ are $\mathcal{T}_{(4)}$, which will be 
integrated along the 4-chain produced by the deformation of the 3-cycle $\Pi_3$, and a component ${\cal T}_{2}$ integrated over the 2-chain spanned by the 1-cycle in such deformation. The latter accounts for the contribution to the superpotential of the induced D4-brane charge arising from possible D6-brane worldvolume fluxes on the 2-forms Poincar\'e dual to the 1-cycle in $\Pi_3$.

The resulting variation of the superpotential is
\begin{equation}
\Delta W_{D6} = \int_{\Sigma_4} \mathcal{T}_{(4)} + \int_{\Sigma_4} \mathcal{T}_{(2)} \wedge \mathcal{F} = \int_{\Sigma_4} \mathcal{T}_{(4)} + \int_{\Sigma_2} \mathcal{T}_{(2)}, 
\end{equation}
where $\mathcal{F}$ is the magnetic field induced in the D6-brane.

 Therefore, the only way to reproduce the D6-brane open string moduli dependence described by the Ganor zeroes is that the instanton backreaction indeed produces the deformation $\delta {\cal T}$ with components ${\cal T}_{(2)}$ and ${\cal T}_{(4)}$.

\section{Gauge non-perturbative effects}
\label{sec:gauge}

There are D3-brane instantons which admit the interpretation of gauge theory instantons. This happens when the D3-brane instanton wraps precisely the same 4-cycle as a stack of  4d spacetime-filling D7-branes. 

The description of stabilization of axions coupling to non-abelian gauge interactions has been recast in terms of coupling to the composite Chern-Simons 3-form in \cite{Dvali:2005an,Dvali:2005ws,Dvali:2013cpa} c.f. section \ref{sec:review}. It is natural to ask about any possible interplay between this and the 3-form discussed in earlier sections.

Actually, the 3-form in earlier sections arises only when the D-brane instantons backreact, in other words when the gauge dynamics is geometrized. This implies that we must consider the geometry that results when the D7-branes, together with the euclidean D3-brane instantons, are backreacted on the geometry. The resulting configuration no longer contains open string degrees of freedom, as everything is encoded in the backreacted geometry. Therefore the relation between the 3-forms is essentially holographic: on one side there are open string degrees of freedom, and the axion stabilization mechanism is described as in \cite{Dvali:2005an,Dvali:2005ws,Dvali:2013cpa} in terms of a 3-form constructed out of the open string sector gauge fields; on the other side, there is a backreacted geometry, and no open string degrees of freedom, and the axion stabilization arises from a 3-form supported by the distorted geometry.

To flesh out the latter claim, let us consider the description of the backreaction of D7-branes and their euclidean D3-brane instanton effects. We concentrate in the case ${\cal N}=1$ SYM, where the instantons contribute to the superpotential, and therefore to the axion stabilization.

To describe the backreacted geometry, we borrow results from \cite{Dymarsky:2010mf}. The backreaction of the D7-branes, at the perturbative level (i.e not including the euclidean D3-brane instantons) is defined by
\beqa
{\cal Z}\, =\, \Omega \quad ,\quad {\cal T}\, =\, e^{-\Phi} \exp  ( i e^{\Phi/2} J)
\eeqa
(where here $\Phi$ is the dilaton).

To this order there is no deformation of ${\cal Z}$, and moreover no 1-form that can support the 3-form. The latter is however generated when the non-perturbative gaugino condensate (which is described by (fractional) euclidean D3-branes) is included. The gaugino condensate $\left\langle S \right\rangle = \left\langle \lambda \lambda \right\rangle = e^{2 \pi i k/n} \Lambda^3$ is a non-zero vev for the gaugino bilinear. In  \cite{Dymarsky:2010mf}  it was found that
\beqa
d{\cal Z}\,=\, i\ell_s^4 \langle S\rangle \delta_2(\Sigma)
\eeqa
This is exactly as in (\ref{magic}), thereby confirming the anticipation that the backreacted D7/D3 system can support a 3-form in agreement with the mechanism in earlier sections.

\section{Conclusions}
\label{sec:conclusions}

In this note we have provided the description of the axion potential from non-gauge D-brane instanton effects in terms of a 3-form eating up the  2-form dual to the axion. The 3-form arises from the KK reduction of higher-dimensional RR fields in the generalized geometry arising when the D-brane instanton backreaction is taken into account. The mechanism also holds for D-brane instantons corresponding to gauge instantons, in which case the generalized geometry description of the axion couplings can be regarded
as holographically related to earlier 3-form descriptions of stabilization of QCD-like axions.

Our works puts axion potentials from non-perturbative effects in a similar footing to other stabilization mechanisms, like flux compactifications. We hope this can improve the study of the interplay between different stabilization mechanisms in string theory. In another line,  given recent  results in applying the Weak Gravity Conjecture to axion models  in terms of their dual 3-forms \cite{Ibanez:2015fcv}, we expect our analysis to allow similar analysis for non-perturbative axion potentials from instantons.

We hope to come back to these questions in the near future.

\bigskip

%=======================================================
\section*{Acknowledgements}
%=======================================================

We would like to thank L. Ib\'a\~nez, F. Marchesano, A. Retolaza and G. Zoccarato for useful discussions. E. G. and A. U. are partially supported by the grants FPA2015-65480-P from the MINECO, the ERC Advanced Grant SPLE under contract ERC-2012-ADG-20120216-320421 and the grant SEV-2012-0249 of the ``Centro de Excelencia Severo Ochoa" Programme.

\newpage

\bibliographystyle{JHEP}
\bibliography{mybib}

\end{document}